\documentclass[prr,aps,twocolumn,10pt,superscriptaddress,notitlepage,nofootinbib,longbibliography]{revtex4-2}
\usepackage[margin=0.75in]{geometry}
\usepackage[caption=false]{subfig}
\usepackage{graphicx}
\usepackage{amsmath,mathtools,bm}
\usepackage{amssymb}
\usepackage{epstopdf}
\usepackage{siunitx}
\usepackage[version=4]{mhchem}
\usepackage{color}
\usepackage[ngerman,english]{babel}

\usepackage{lipsum, babel}

\usepackage{tabularx}

\usepackage{braket}

\definecolor{winered}{rgb}{0.5,0,0}

\usepackage{letltxmacro}
\LetLtxMacro{\ORIGselectlanguage}{\selectlanguage}
\makeatletter
\DeclareRobustCommand{\selectlanguage}[1]{%
  \@ifundefined{alias@\string#1}
    {\ORIGselectlanguage{#1}}
    {\begingroup\edef\x{\endgroup
       \noexpand\ORIGselectlanguage{\@nameuse{alias@#1}}}\x}%
}
\newcommand{\definelanguagealias}[2]{%
  \@namedef{alias@#1}{#2}%
}

\DeclareSymbolFont{matha}{OML}{txmi}{m}{it}
\DeclareMathSymbol{\varv}{\mathord}{matha}{118}

\usepackage{amsmath}

\usepackage{graphicx,wrapfig}

\usepackage{amsmath}
\usepackage{float}
\makeatother

\definelanguagealias{en}{english}
\definelanguagealias{EN}{english}
\definelanguagealias{eng}{english}
\definelanguagealias{de}{ngerman}
\usepackage[utf8]{inputenc}

\usepackage[colorlinks]{hyperref}
 \hypersetup{
 colorlinks=true,
 linkcolor=blue,
 urlcolor={blue},
 filecolor={blue},
 citecolor={blue},
 allcolors={blue}
 }

\usepackage{bibunits}
\defaultbibliographystyle{apsrev4-2}
\defaultbibliography{DiamondOMCs}

\begin{document}

\raggedbottom
\begin{bibunit}

\title{Exceptional points in diamond optomechanics}

\author{Waleed El-Sayed}
\affiliation{Institute for Quantum Science and Technology, University of Calgary, Calgary, AB, T2N 1N4, Canada}
\author{Elham Zohari}
\affiliation{Institute for Quantum Science and Technology, University of Calgary, Calgary, AB, T2N 1N4, Canada}
\affiliation{Department of Physics, University of Alberta, Edmonton, AB, T6G 2E1, Canada}
\affiliation{National Research Council of Canada, Quantum and Nanotechnology Research Centre, Edmonton, Alberta, T6G 2M9,
Canada}
\author{Joe Itoi}
\affiliation{Institute for Quantum Science and Technology, University of Calgary, Calgary, AB, T2N 1N4, Canada}
\author{Peyman Parsa}
\affiliation{Institute for Quantum Science and Technology, University of Calgary, Calgary, AB, T2N 1N4, Canada}
\author{Gustavo de Oliveira Luiz}
\affiliation{Institute for Quantum Science and Technology, University of Calgary, Calgary, AB, T2N 1N4, Canada}
\affiliation{nanoFAB Centre, University of Alberta, Edmonton, AB, T6G 2V4, Canada}
\author{Joseph E. Losby}
\affiliation{Institute for Quantum Science and Technology, University of Calgary, Calgary, AB, T2N 1N4, Canada}
\author{Misa Hayashida}
\affiliation{National Research Council of Canada, Quantum and Nanotechnology Research Centre, Edmonton, Alberta, T6G 2M9,
Canada}
\author{Marek Malac}
\affiliation{National Research Council of Canada, Quantum and Nanotechnology Research Centre, Edmonton, Alberta, T6G 2M9,
Canada}
\author{Paul E. Barclay}
\email[Paul~E.\ Barclay: ]{Corresponding author pbarclay@ucalgary.ca}
\affiliation{Institute for Quantum Science and Technology, University of Calgary, Calgary, AB, T2N 1N4, Canada}

\date{\today}

\begin{abstract}

Multimode cavity optomechanical systems allow light to couple otherwise non-interacting mechanical resonators, enabling non-Hermitian phenomena such as exceptional points, where eigenfrequencies and eigenvectors of coupled modes coalesce. Accessing an exceptional point and its nearby parameter space is a first step towards chiral mode dynamics and topological state transfer. Diamond optomechanical devices support strong coherent optomechanical coupling required to tune resonances to an exceptional point, as well as strain-coupling to spin-defects for hybrid quantum technologies, but have not yet been used for multimode non-Hermitian physics. Here we tune to an exceptional point in a diamond optomechanical crystal, which uses structural symmetry breaking to produce two high-frequency mechanical resonances coupled to an optical cavity. The exceptional point is reached within a stable operating window below the phonon-lasing threshold, and we observe asymmetric redistribution of optomechanical damping and anti-damping between hybridized modes. These results establish diamond optomechanical crystals as a platform for non-Hermitian optomechanics, opening routes to topological mechanical dynamics in hybrid spin-phonon interfaces. 
\end{abstract}

\maketitle



\begin{figure*}[th!]
    \includegraphics[width=1\linewidth]{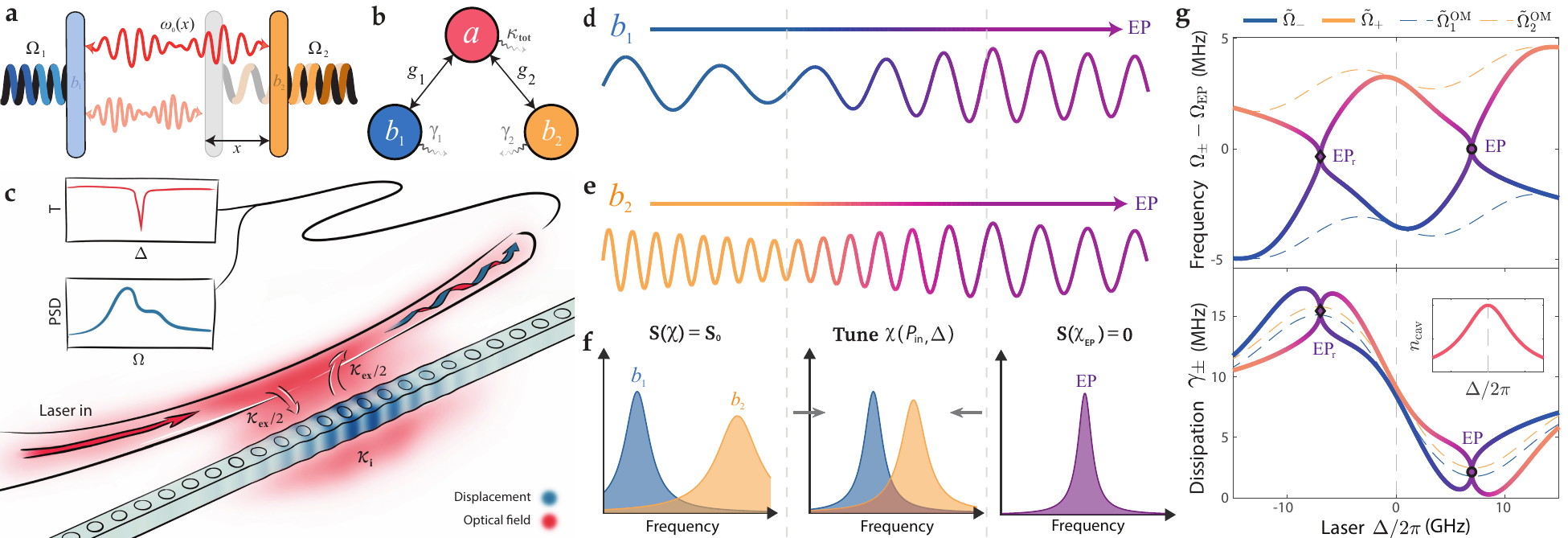}
		\caption{ \small \textbf{Exceptional points in multimode cavity optomechanics.} 
        (\textbf{a}) Fabry--Perot representation of a multimode optomechanical system, where two mechanical resonators with frequencies $\Omega_{1,2}$ modulate a common optical resonance $\omega_c$. 
        (\textbf{b}) Coupling diagram for two mechanical modes $\hat{b}_1$ and $\hat{b}_2$ coupled to an optical mode $\hat{a}$ with vacuum optomechanical coupling rates $g_1$ and $g_2$. The mechanical dissipation rates are $\gamma_1$ and $\gamma_2$, and the optical decay rate is $\kappa=\kappa_{\rm ex}+\kappa_{\rm i}$.
        (\textbf{c}) Diamond optomechanical crystal (OMC) implementation. A fiber taper evanescently couples laser light into and out of the optical cavity, allowing the thermomechanical spectrum to be measured from the transmitted field.
        (\textbf{d},\textbf{e}) Conceptual time-domain evolution of two mechanical modes as optically mediated coupling tunes the system toward an exceptional point (EP), where the frequencies and decay rates coalesce.
        (\textbf{f}) Corresponding spectral evolution. Tuning the optical susceptibility $\chi(\Delta,P_{\rm in})$ toward $\chi_{\rm EP}$ merges the two mechanical resonances when the complex splitting $S$ vanishes. 
        (\textbf{g}) Calculated complex eigenfrequencies $\tilde{\Omega}_{\pm}=\Omega_{\pm}-i\gamma_{\pm}/2$ versus laser detuning $\Delta$ for a representative two-mode system with $g_1=g_2$ and $|\Omega_1-\Omega_2|/2\pi=7~\mathrm{MHz}$. Blue- and red-detuned EPs occur on opposite sides of the optical resonance. Dashed curves show $\tilde{\Omega}^{\rm OM}_j$, the optical spring and damping response of each mode without the optically mediated off-diagonal coupling $g_1g_2\chi$ between them.
        }
            \label{fig:Fig1}
\end{figure*}

Nanoscale mechanical resonators can interact with a broad range of quantum systems, including superconducting circuits \cite{Xiang2013}, solid-state spins~\cite{Lee-SpinMechanicsReview-JoO-2017, wang-SpinMechanicsCoupling-2020-arxiv}, and photons~\cite{Janitz2020}, making them powerful intermediaries for hybrid quantum technologies~\cite{barzanjeh2021optomechanics, bowen2015_quantum}. When combined with optical cavities, they can be coherently controlled using light, enabling classical and quantum information processing and sensing technologies \cite{aspelmeyer_cavity_2014_rev, barzanjeh2021optomechanics}. Many of the technologies being advanced by such cavity optomechanical systems, such as directional phonon transport and isolation~\cite{xu2016_topologicalET, Xu2019Nonreciprocal, Ren2022TopologicalPhononTransport}, topological state transfer between mechanical modes~\cite{Guria2024ResolvingTopologyMultipleEPs, Lai2024NonreciprocalTopologicalPhononTransfer}, directional amplification for quantum-limited readout~\cite{delPino2022NonHermitianChiralPhononics, Slim2024BosonicKitaevChain}, and tunable spin-strain coupling to spin defects~\cite{Lee2025TuningStrainCoupling}, are inherently multimode and require non-Hermitian engineering of coupled mechanical resonances. A key element in these systems is exceptional points (EPs): spectral singularities where the eigenvalues and eigenvectors of coupled modes coalesce~\cite{Heiss2012JPhysAEP, Miri2019ScienceEPReview}. Dynamically encircling an EP in parameter space yields chiral, direction-dependent energy and state transfer~\cite{ElGanainy2018NatPhysNHReview, Zhong2018WindingNonHermitianSingularities, Doppler2016DynamicallyEncirclingEP, Guria2024ResolvingTopologyMultipleEPs, xu2016_topologicalET}, while the enhanced spectral response near an EP has been investigated as a mechanism for sensing across photonic and electromechanical platforms~\cite{Mao2024EPPhaseSensing,Kononchuk_2022}. Operating near an EP also enables strong, asymmetric gain redistribution between coupled modes via eigenvector hybridization ~\cite{xu2016_topologicalET,ElGanainy2018NatPhysNHReview}.  

Realizing an EP with mechanical resonators is possible by establishing tunable coupling between them through their interaction with a common optical cavity field, which can modify their frequency and damping \cite{Shkarin2014, Lin2010NatPhotonCoherentMixing}. 
Mechanical EPs in cavity optomechanics have been demonstrated with membrane resonators~\cite{xu2016_topologicalET, delPino2022NonHermitianChiralPhononics} and with a pair of nearly identical optically coupled silicon optomechanical crystals (OMCs) \cite{wu2023_onchipEP}. 
Here we realize an EP with two mechanical resonances localized within a single OMC fabricated from single-crystal diamond. Diamond's physical properties, which allow it to support intense optical fields required to induce the strong coupling needed for coalescence, combined with its ability to host spin defects such as nitrogen-vacancy (NV) and silicon-vacancy (SiV) color centers, make it an attractive platform for building cavity optomechanical systems with EPs.  In particular, by coupling the electronic states of spin-defects to mechanical strain~\cite{shandilya2022_diamond, Teissier2014PRLStrainNVDiamond,  Ovartchaiyapong2014NatCommunStrainNV, MacQuarrie2013PRLMechSpinControl, maity2020coherent}, OMCs may enable on-chip spin-phonon-photon interfaces~\cite{joe2024_highq_diamondomc_siv, Oh2026SpinEmbeddedDiamondOMR, Joe2026PurcellEnhancedSpinPhonon}. Incorporating multimode phenomena such as EPs and related topological effects will further extend the versatility of diamond optomechanics platforms.

Using our diamond OMC, we systematically tune the effective coupling between two mechanical modes through the EP regime. This occurs within a stable operating regime before the onset of phonon lasing~\cite{Kippenberg2005PRLRadPressureOsc, Rokhsari-2005-OpticsExpress-SelfOscillations, krause2015_nonlinear}, and we show that the non-Hermitian coupling redistributes dynamical backaction asymmetrically between mechanical modes, enhancing the optical driving of mechanical modes possessing weaker optomechanical coupling. Furthermore, in our system the OMC's multimode spectrum is created through vertical symmetry breaking that mixes cavity-localized mechanical breathing mode with a flexural mode in the optical mirror region. The resulting extended mechanical mode profiles can couple to spins embedded in the diamond far from the optical driving field, shielding them from decoherence effects that can limit the performance of spin-optomechanical devices \cite{joe2024_highq_diamondomc_siv}. Moreover, tuning the optically mediated hybridization of the two modes provides a path to dynamically tune the interference between their local strain profiles, enabling in situ control of the relative optomechanical and spin--strain coupling strengths. 


\subsection*{Multimode optomechanics}

The canonical multimode cavity optomechanical system used to realize an EP is illustrated in Figs.~\ref{fig:Fig1}\textbf{a,b}, and is described by Hamiltonian~\cite{aspelmeyer_cavity_2014_rev}
\begin{equation}
\label{eq:H}
\hat{H} = \hbar\omega_{c}\hat{a}^{\dagger}\hat{a} + \sum_{j=1,2}\hbar\Omega_{j}\hat{b}_{j}^{\dagger}\hat{b}_{j} - \sum_{j} \hbar g_{j} \hat{a}^{\dagger} \hat{a}\left(\hat{b}_{j}^{\dagger}+\hat{b}_{j}\right),
\end{equation}
where  $\omega_{\rm c}$ and $\Omega_{j}$ (for $j=1,2$) are the optical and mechanical resonance frequencies, $\hat{a}$ ($\hat{a}^{\dagger}$) and $\hat{b}_j$ ($\hat{b}^{\dagger}_j$) are photon and phonon annihilation (creation) operators, respectively, and the vacuum optomechanical coupling rates $g_{j}$ set the optical frequency shift per zero-point displacement of each mechanical mode. 
The finite response time of the driven optical cavity field relative to the mechanical motion produces dynamical backaction: a shift in each mechanical resonance frequency (optical spring effect) and a modification of its damping rate (optical damping or amplification), captured by the real and imaginary parts of $g_j^2\chi$, respectively, where $\chi$ is the complex susceptibility introduced below. When multiple mechanical modes couple to the same optical resonance, this backaction also generates an effective coupling between them, providing access to the rich topological landscape surrounding EPs where the frequencies, damping rates, and eigenvectors of the coupled modes coalesce~\cite{Heiss2012JPhysAEP, Miri2019ScienceEPReview}. 

\begin{figure*}[t]
    \includegraphics[width=1.0\linewidth]{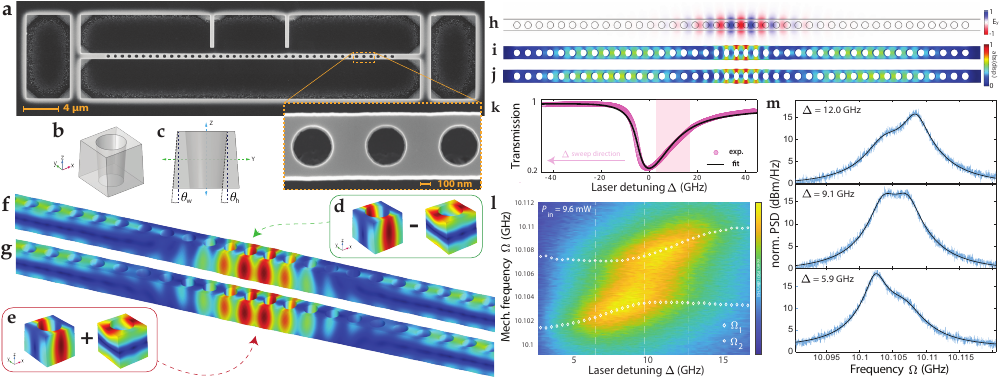}
		\caption{
		    \small \textbf{Symmetry-mixed mechanical modes in a diamond optomechanical crystal.}
            (\textbf{a}) Scanning electron micrograph of the fabricated diamond optomechanical crystal (OMC) nanobeam.
            (\textbf{b},\textbf{c}) Schematic unit cell showing the sidewall angles $\theta_w$ and $\theta_h$, which break vertical reflection symmetry.
            (\textbf{d},\textbf{e}) Mode-composition diagrams showing a breathing-like cavity unit-cell mode combined with a flexural-like mirror unit-cell mode. The minus and plus signs denote the relative displacement phase of the two parent mode components, giving the out-of-phase and in-phase mixed modes shown in (\textbf{f}) and (\textbf{g}), respectively.
            (\textbf{f},\textbf{g}) Simulated displacement profiles of the two symmetry-mixed mechanical modes in the full OMC cavity. Both retain displacement in the optical-cavity region, enabling coupling to the same optical mode.
            (\textbf{h}) Simulated $E_y$ field of the fundamental TE-like optical cavity mode.
            (\textbf{i},\textbf{j}) Top-view displacement profiles of the two mechanical modes, showing their spatial extent beyond the optical-cavity region.
            (\textbf{k}) Normalized fiber-taper transmission spectrum across the optical cavity resonance at $P_{\rm in}=9.6$~mW. The solid curve is a phenomenological transmission model using optical parameters obtained from the global optomechanical fit, with a Fano background included to reproduce the measured asymmetry. The shaded region marks the detuning window used for mechanical spectra throughout this work.
            (\textbf{l}) Measured mechanical power spectral density (PSD) versus laser detuning $\Delta$, with white markers showing the fitted eigenfrequencies $\Omega_\pm$.
            (\textbf{m}) Mechanical PSDs at the detunings marked in (\textbf{l}); black curves are coupled-mode fits used to extract $\Omega_\pm$ and $\gamma_\pm$.
            }
            \label{fig:Fig2}
\end{figure*}

This effect is captured by the complex frequencies $\tilde{\Omega}_{\pm}$ of the system's hybridized mechanical modes, predicted from an effective Hamiltonian governing the mechanical amplitudes $i\dot{\bf b} = H_{\rm eff}{\bf b}$~\cite{wu2023_onchipEP,xu2016_topologicalET}:
\begin{equation}
\label{eq:Heff}
H_{\rm eff}=
\begin{pmatrix}
\Omega_{1}-i \gamma_{1} / 2+g_{1}^{2} \chi & g_{1} g_{2} \chi\\[4pt]
g_{1} g_{2} \chi & \Omega_{2}-i \gamma_{2} / 2+g_{2}^{2} \chi
\end{pmatrix}
\end{equation}
%
%
%
where ${\bf b} = [\hat{b}_1,\hat{b}_2]^T$. This expression is derived from Langevin equations that incorporate the optical cavity decay rate, $\kappa$, the intrinsic mechanical dissipation rates $\gamma_j$, and an external laser drive.
Here $\chi=n_{\rm cav}[(\omega+\Delta+i \kappa / 2)^{-1}-(\omega-\Delta+i \kappa/2)^{-1}]$ is the photon-number-weighted difference between the cavity susceptibilities at the anti-Stokes and Stokes sidebands of the drive, with $\omega = (\Omega_1 + \Omega_2)/2$ and $n_{\rm cav}(P_{\rm in},\Delta)$ the intracavity photon number set by the laser input power $P_{\rm in}$ and detuning $\Delta$ from cavity resonance. For a cold cavity, $\Delta \equiv (\omega_L - \omega_{\rm cav})$ (see Supplementary Information \ref{Supp:fitting}). 

Diagonalizing $H_{\rm eff}$ yields the complex eigenfrequencies $\tilde{\Omega}_{\pm} = \Omega_{\pm} - i\gamma_{\pm}/2$:
\begin{equation} \label{eq:EP_eigenfreq}
\tilde{\Omega}_{\pm}=\frac{1}{2}(\tilde{\Omega}_{1}^{\rm OM}+\tilde{\Omega}_{2}^{\rm OM}) \pm \frac{1}{2}S,
\end{equation}
with
\begin{equation} \label{eq:EP_splitting}
S=\sqrt{(\tilde{\Omega}_{1}^{\rm OM}-\tilde{\Omega}_{2}^{\rm OM})^{2}+4(g_{1}g_{2}\chi)^{2}}.
\end{equation}
The diagonal terms $\tilde{\Omega}_{j}^{\rm OM} = \Omega_{j}-\tfrac{i}{2}\gamma_{j}+g_{j}^{2}\chi$ are the single-mode complex frequencies set by the optical spring and damping effects. The off-diagonal terms $g_1g_2\chi$ is the cavity-mediated coupling between the two mechanical modes, which is the key to engineering an EP: when the complex splitting $S$ vanishes, the eigenvalues and eigenvectors coalesce ($\tilde{\Omega}_+ = \tilde{\Omega}_-$).
This condition is satisfied when 
\begin{equation}
    \label{eq:chiEP}
    \chi(P_{\rm in},\Delta) = \chi_{\rm EP}^{(\pm)} \equiv \frac{\Delta\Omega-\tfrac{i}{2}\Delta\gamma}{(g_2^2 - g_1^2) \pm i\,2g_1g_2},
\end{equation}
where $\Delta\Omega = \Omega_1 - \Omega_2$ and $\Delta\gamma = \gamma_1 - \gamma_2$ \cite{wu2023_onchipEP}. The EP therefore appears as a target point $\chi_{\rm EP}$ in the optically controlled susceptibility plane, accessed experimentally by tuning $P_{\rm in}$ and $\Delta$. In our diamond OMC, $P_{\rm in}$ is delivered through a fiber taper evanescently coupled to the cavity (Fig.~\ref{fig:Fig1}{\bf c}).

Figs.~\ref{fig:Fig1}{\bf d--f} illustrate how the mechanical spectrum evolves as $\chi(P_{\rm in},\Delta)$ is tuned to $\chi_{\rm EP}$, and Fig.~\ref{fig:Fig1}{\bf g} quantitatively studies the eigenvalues of a representative system tuned to reach an EP. In this example with balanced optomechanical coupling rates ($g_1 = g_2$), the optical spring effect is identical for both modes, and two EPs appear symmetrically at red- and blue-detuning for the same $n_\text{cav}$. In experimental devices, the coupling symmetry is typically broken ($g_1 \neq g_2$), leading to unique power requirements for each EP. Furthermore, characterizing the red-detuned EP can be hindered by photothermal instability and optomechanical damping. We therefore focus on the blue-detuned EP. Because blue-detuned backaction optomechanically amplifies the mechanical modes, accessing the EP requires that it lie below the phonon-lasing threshold ($\gamma_\pm = 0$), beyond which the linearized analysis breaks down and the EP becomes inaccessible. The resulting constraints on device parameters and the accessible $\chi(P_{\rm in}, \Delta)$ are discussed in Supplementary Information~\ref{supp:stability}.

\subsection*{Multimode Diamond OMC}
The OMC studied here is formed from a diamond nanobeam patterned with a photonic and phononic crystal lattice. Gradually transitioning the nominal periodic waveguide unit cell towards a cavity defect in the center of the nanobeam creates spatially overlapping optical and mechanical modes. During fabrication (see Supplementary Information~\ref{supp:fab}), the diamond etching process introduces small sidewall angles, breaking the device's vertical mirror symmetry, as illustrated in Figs.~\ref{fig:Fig2}{\bf b--c}. This symmetry breaking mixes mechanical modes that would otherwise be orthogonal, allowing a breathing mode in the cavity region and a nearly degenerate flexural-like mode in the waveguide region to combine to form two extended mechanical modes with in-phase and out-of-phase motion, whose displacement profiles are shown in Figs.~\ref{fig:Fig2}{\bf d--g}. Both of these symmetry-mixed modes have displacement in the center of the nanobeam where the device's optical cavity mode is confined, as shown in Fig.~\ref{fig:Fig2}{\bf h}, and therefore couple strongly to it. Throughout this work, $\hat{b}_1$ and $\hat{b}_2$ denote these mixed extended mechanical modes of the fabricated OMC. We reserve the terms hybridization and hybridized modes for the optically dressed eigenmodes that diagonalize $H_{\rm eff}$. 
\subsection*{Device Characterization}
The device was characterized using a dimpled fiber taper to couple tunable laser light into and out of the diamond OMC. The measured transmission of the fundamental TE-like cavity mode ($\lambda_0 = 1565.04$ nm) is shown in Fig.~\ref{fig:Fig2}\textbf{k}. Because thermo-optic distortion and Fano interference complicate direct linewidth extraction from this trace alone, the optical parameters used below were obtained from a global fit to the detuning- and power-dependent mechanical response, with the ratio of extrinsic fiber-coupling rate to total decay rate, $\kappa_{\rm ex}/\kappa$, constrained by the measured transmission dip. This gives $\kappa_{\rm ex}/2\pi = 7.03~\mathrm{GHz}$ and $\kappa/2\pi = 15.83~\mathrm{GHz}$.

The mechanical response was measured from the power spectral density (PSD) of the transmitted laser intensity versus detuning $\Delta$ for a given input power $P_{\rm in}$. The spectrogram in Fig.~\ref{fig:Fig2}\textbf{l} shows two thermally driven mechanical resonances whose frequencies and relative amplitudes change with detuning, consistent with optical spring shifts and optically mediated intermode coupling. Representative PSD slices are shown in Fig.~\ref{fig:Fig2}\textbf{m}. We extract the complex eigenfrequencies $\tilde{\Omega}_\pm(\Delta)$ by fitting the spectra to a coupled-mode line shape that accounts for the non-Lorentzian multimode response (Supplementary Information~\ref{Supp:fitting}). Fitting $\tilde{\Omega}_\pm(\Delta)$ to Eq.~\ref{eq:EP_eigenfreq} yields $\Omega_1/2\pi = 10.1016~\mathrm{GHz}$, $\Omega_2/2\pi = 10.1085~\mathrm{GHz}$, $\gamma_1/2\pi = 8.45~\mathrm{MHz}$, $\gamma_2/2\pi = 9.14~\mathrm{MHz}$, $g_1/2\pi = 298.6~\mathrm{kHz}$ and $g_2/2\pi = 285.2~\mathrm{kHz}$. We attribute the discrepancy between the fitted and simulated optomechanical coupling rates ($g_{1,\rm sim}/2\pi = 262~\mathrm{kHz}$, $g_{2,\rm sim}/2\pi = 259~\mathrm{kHz}$) primarily to the fiber-taper loss estimate used to determine $n_{\rm cav}$.

\subsection*{Experimentally Tuning across a Mechanical Exceptional Point}
\begin{figure*}[!t]
	\includegraphics[width=1\linewidth]{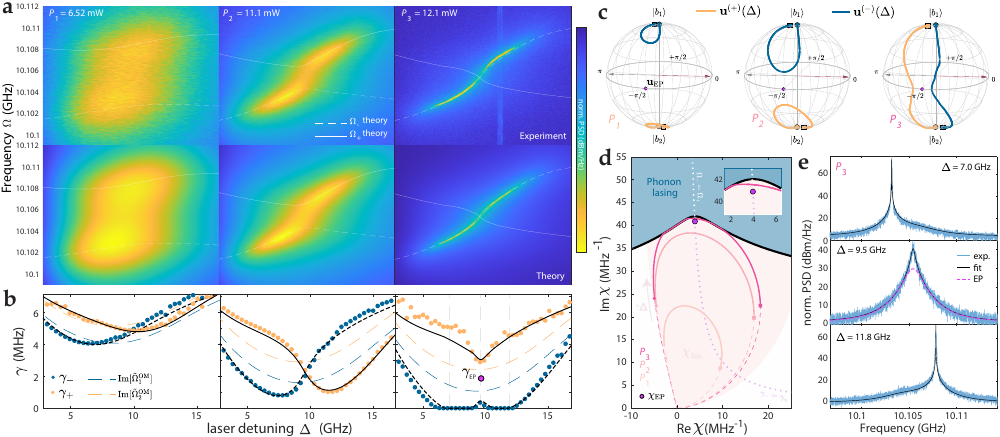}
		\caption{
		    \small {\bf Optomechanical tuning through the exceptional-point regime.} (\textbf{a}) Mechanical power spectral density (PSD) versus laser detuning $\Delta$ for three input powers $P_1$, $P_2$ and $P_3$. Top row, measured spectra; bottom row, spectra calculated from the fitted two-mode model. White curves show the predicted eigenfrequencies $\Omega_{\pm}$. Increasing input power strengthens the optically mediated coupling, drawing the frequency branches together and modifying their damping.
            (\textbf{b}) Extracted damping rates $\gamma_{\pm}$ (points) and model predictions (black curves) versus $\Delta$. Colored dashed curves show the damping expected from the diagonal optical backaction terms alone, excluding the off-diagonal coupling $g_1g_2\chi$. At $P_3$, the frequency branches have crossed while the fitted damping rates remain unequal, placing the system beyond the inferred EP. The purple marker indicates the inferred EP damping rate; dashed vertical lines mark the detunings used in (\textbf{e}).
            (\textbf{c}) Bloch-sphere trajectories of the eigenvectors $u^{(\pm)}(\Delta)$, with coordinates $z=|u_1|^2-|u_2|^2$ and $\phi=\arg(u_2/u_1)$. Red and gray arrows denote the bright and dark superpositions that maximize and minimize $g_{\rm eff}=|\mathbf{g}\cdot\mathbf{u}|$, respectively; the purple point marks the EP eigenvector.
            (\textbf{d}) Susceptibility trajectories $\chi(P_{\rm in}, \Delta)$ in the complex plane. The black contour marks the instability threshold $\gamma_{\pm}=0$. Dotted curves mark the contours $\Omega_+=\Omega_-$ and $\gamma_+=\gamma_-$, which meet at the EP condition $\chi_{\rm EP}$. At $P_3$, increasing $\Delta$ first drives the system across the instability threshold, then through a stable-regime detuning window beyond the inferred EP, before phonon-lasing returns at larger detuning.
           (\textbf{e}) Measured PSD slices at $P_3$ for the detunings marked in (\textbf{b}). The upper and lower traces show phonon lasing for $\Delta$ outside the stability window. The middle trace lies inside this window and appears as a single resonance; the coupled-mode line-shape fit assigns it to a post-EP state with coalesced real frequencies and unequal damping rates. The purple dashed curve shows the inferred EP response calculated from the fitted parameters, where $\Omega_+=\Omega_-$ and $\gamma_+=\gamma_-$.
            }
            \label{fig:Fig3}
\end{figure*}

\begin{figure*}[!t]
	\includegraphics[width=0.9\linewidth]{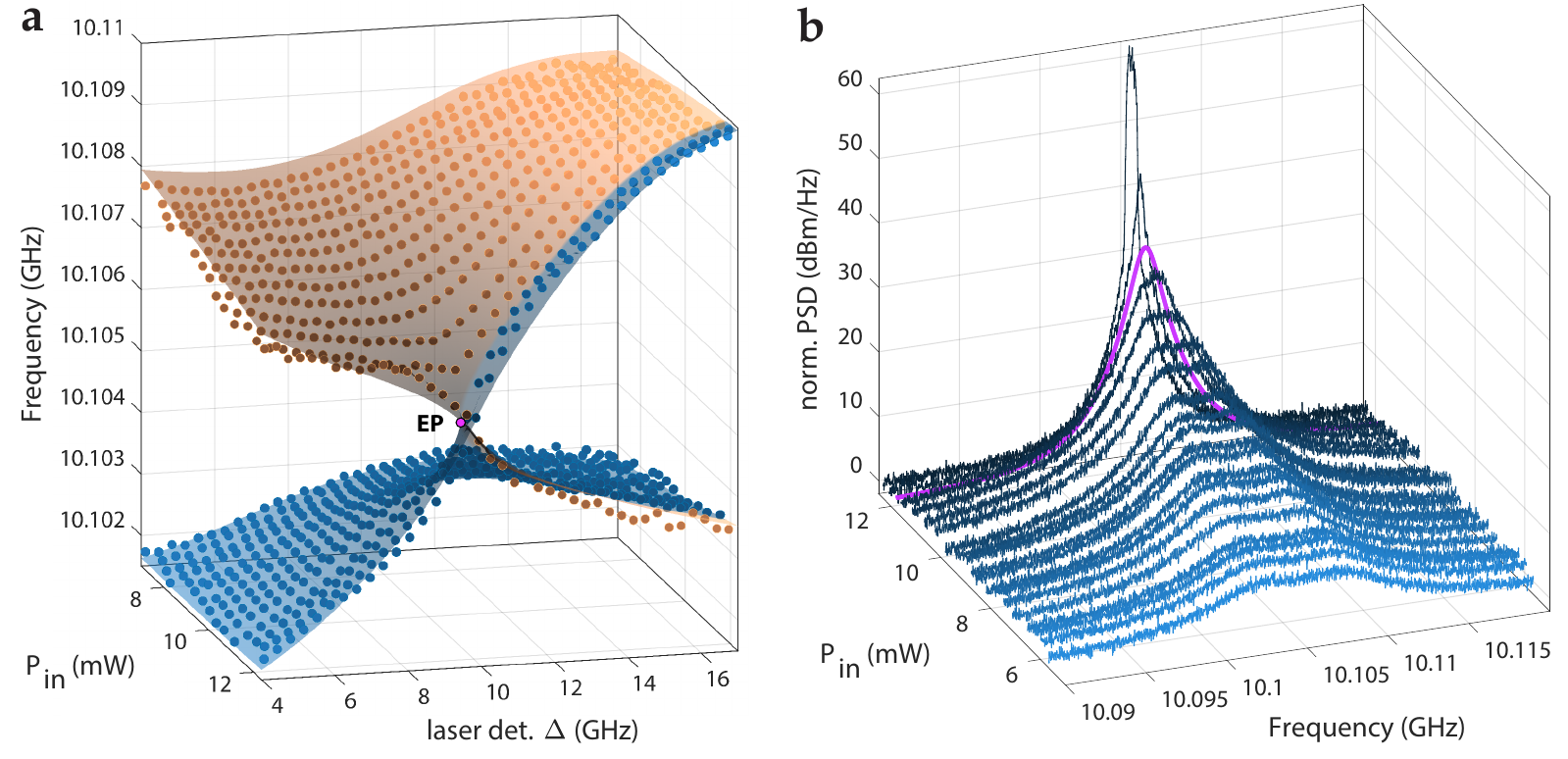}\centering
		\caption{
		    \small \textbf{Branch-point topology of the mechanical exceptional point.} (\textbf{a}) Measured real eigenfrequencies $\Omega_\pm$ (markers) and fitted two-mode model surface versus laser detuning $\Delta$ and input power $P_{\rm in}$. The two frequency sheets approach and meet at the inferred EP (purple point). Beyond the EP, the continuously evolved eigenbranches reconnect onto the opposite frequency sheets, with the $(\pm)$ labels following branch continuity rather than frequency ordering.
            (\textbf{b}) Mechanical power spectral densities (PSDs) along the two degeneracy contours shown in Fig.~\ref{fig:Fig3}\textbf{d}. For $P_{\rm in}<P_{\rm EP}$, spectra follow the damping-degeneracy contour $\gamma_+=\gamma_-$, isolating the collapse of the real-frequency splitting. For $P_{\rm in}>P_{\rm EP}$, spectra follow the frequency-degeneracy contour $\Omega_+=\Omega_-$, showing the beyond-EP evolution with coalesced real frequencies and separating fitted damping rates. The purple trace shows the inferred EP response calculated from the fitted model, where $\Omega_+=\Omega_-$ and $\gamma_+=\gamma_-$ simultaneously. Phonon lasing appears at the highest power shown.
        }
        \label{fig:Fig4}
\end{figure*}

To access the exceptional-point regime in the diamond OMC, the mechanical response is measured while scanning laser detuning over a range of input powers $P_{\rm in}$. Fig.~\ref{fig:Fig3}\textbf{a} shows three representative powers that capture the approach to, and passage beyond, the inferred EP. At low power ($P_{1} = 6.52~\mathrm{mW}$), the optically mediated coupling is weak: the resonances remain spectrally resolved, and their detuning-dependent shifts are dominated by the diagonal optical spring terms ${\tilde{\Omega}}_{1,2}^{\rm OM}$.

At $P_{2} = 11.1~\mathrm{mW}$, the off-diagonal coupling $g_1g_2\chi$ becomes large enough to visibly attract the eigenfrequency branches (Fig.~\ref{fig:Fig3}\textbf{a}). Using the fitted two-mode model to calculate the transduced PSD reproduces the measured spectra, capturing the observed evolution of the peak separation, linewidth, and relative spectral weight. This agreement extends to the extracted damping rates $\gamma_\pm(\Delta)$, shown together with the calculated curves in Fig.~\ref{fig:Fig3}\textbf{b}. These damping rates reveal an important signature of non-Hermitian hybridization: a growing separation between $\gamma_+$ and $\gamma_-$.

While optomechanical damping in single-mode systems depends linearly on power ($\gamma_{\rm opt} = -2{\rm Im}[g^2\chi] \propto P_{\rm in}$), the hybridized mechanical modes develop opposing linewidth shifts---one branch narrows (becomes amplified) while the other broadens. This asymmetric backaction, in which optical gain is not distributed equally among eigenfrequency branches, is evident from Eq.~\ref{eq:EP_eigenfreq} where $|\gamma_+ - \gamma_-| =  2|{\rm Im} S|$. This reflects the unequal projection of the two eigenvectors $\mathbf{u}^{(\pm)} = (u_1^{(\pm)},u_2^{(\pm)})^T$ on the coupling vector $\mathbf{g} = (g_1, g_2)^T$:
\begin{equation}
    \label{eq:g_eff}
    g_{\rm eff}^{(\pm)} = |\mathbf{g} \cdot \mathbf{u}^{(\pm)}| =  |g_1 u_1^{(\pm)} + g_2 u_2^{(\pm)}|,
\end{equation}
where the normalized complex coefficients $u_j^\pm(\chi)$ denote the contribution from each bare mode $\hat b_j$. As $\chi$ is tuned, one eigenmode is rotated into stronger alignment with $\mathbf{g}$, enhancing its optomechanical coupling, while simultaneously rotating the other eigenmode out of alignment and suppressing its coupling. This eigenvector rotation is visualized in Fig.~\ref{fig:Fig3}\textbf{c}, where the Bloch-sphere trajectories show the changing modal weights and relative phase of $u^{(+)}$ and $u^{(-)}$ as they approach and pass the EP eigenstate. Their distance from the optically bright and dark directions, $\mathbf{u}_{\rm B}\propto(g_1,g_2)^T$ and $\mathbf{u}_{\rm D}\propto(g_2,-g_1)^T$, with  $g_{\rm eff,B}=\sqrt{g_1^2+g_2^2}$ and $g_{\rm eff,D}=0$, illustrates the redistribution of $g_{\rm eff}$ between branches. In the linear model, increasing $|\chi|$ can rotate the eigenvectors further toward this bright--dark limit, but the corresponding largest-$g_{\rm eff}$ portions of the blue-detuned trajectory lie in the unstable region. Experimentally, the largest accessible redistribution therefore occurs near the phonon-lasing threshold at the highest measured power; at $P_{\rm in}=18.6$~mW and $\Delta/2\pi=18.4~\mathrm{GHz}$, the fitted eigenvectors give $(g_{\rm eff}^{(+)},g_{\rm eff}^{(-)})/2\pi=(177,372)~\mathrm{kHz}$.

At $P_{3} = 12.1~\mathrm{mW}$, the system enters the regime just beyond the EP. Importantly, because the power sweep is discrete, we do not directly measure the system at $\chi_{\rm EP}$. Instead, $P_2$ and $P_3$ bracket the branch-point singularity, with the fitted model placing the EP at $P_{EP} = 11.8~\mathrm{mW}$ and $\Delta_{\rm EP}/2\pi = 9.6~\mathrm{GHz}$. At $P_3$, the real frequencies $\Omega_\pm$ cross while the linewidths $\gamma_\pm$ remain split (Figs.~\ref{fig:Fig3}\textbf{a-b}). Correspondingly, the two eigenbranches exchange identities across the crossing; the branch labels $(\pm)$ are assigned by continuous evolution rather than by frequency ordering alone. This branch switching is the expected topological signature of passing beyond the EP in parameter space.

This behavior is further clarified in Fig.~\ref{fig:Fig3}\textbf{d}, which maps the trajectories of $\chi(P_{\rm in},\Delta)$ in the complex plane. As $P_{\text{in}}$ increases, the susceptibility loop expands towards the instability threshold defined by $\gamma_{\pm} = 0$. Above this boundary, radiation-pressure-induced self-oscillation (phonon lasing) occurs and nonlinear gain clamping prevents the system from moving deeper into the unstable region. The trajectory in the $\chi$-plane is then pinned to the threshold contour and evolves along it as $\Delta$ is further swept (see Supplementary Information~\ref{supp:stability}) ~\cite{wu2023_onchipEP}. At $P_{3}$, this boundary is reached at detunings where hybridization redistributes the optical backaction unevenly between the two eigenbranches, enhancing the effective optomechanical coupling $g_{\rm eff}$ of one eigenbranch and thereby lowering the lasing threshold. Near $\chi_{\rm EP}$, however, this redistribution no longer favors one branch: at the EP itself, $u_{1, \rm EP} = \pm i u_{2, \rm EP}$ and the coalesced eigenstates have the same projected coupling, $g_{\rm eff, \rm EP} = \frac{1}{\sqrt{2}}|g_1 \pm i g_2| = \sqrt{{(g_1^2 + g_2^2)}/{2}}= 292~\mathrm{kHz} \times 2\pi$. In our system, this creates a narrow detuning window around the EP condition within which the system can remain stable, even after passing to the far side of the EP transition. The experimental realization of this stability buffer surrounding the EP in parameter space is therefore a central result: it shows that the EP topology remains experimentally accessible over a finite region of parameter space, providing an operating margin required for future EP-encirclement experiments.

Fig.~\ref{fig:Fig3}\textbf{e} shows this through the measured PSD at $P_3$ for the three detunings marked in Fig.~\ref{fig:Fig3}\textbf{b}. At $\Delta{/2\pi}=7.0~\mathrm{GHz}$ and $\Delta/2\pi=11.8~\mathrm{GHz}$, outside the stable detuning window, the spectra show phonon lasing on the eigenbranch with the larger projected optomechanical coupling: $(g^{(+)}_\mathrm{eff},g^{(-)}_\mathrm{eff})/2\pi=(242,355)~\mathrm{kHz}$ and $(235,340)~\mathrm{kHz}$ at the lower and upper detuning, respectively. Because the eigenbranches exchange identities across the EP, the larger-$g_\mathrm{eff}$ branch is the same continuously evolved hybrid mode on both sides of the stable window.

At $\Delta/2\pi=9.5 ~\mathrm{GHz}$, inside the stable window, the measured PSD appears as a single resonance. Fitting this trace with the coupled-mode PSD model gives coalesced real eigenfrequencies but unequal damping rates, with the corresponding $\gamma_\pm$ values shown at the middle vertical line in Fig.~\ref{fig:Fig3}\textbf{b}. To distinguish this post-EP response from the EP itself, we overlay the model response evaluated at the inferred EP using the globally fitted system parameters. At this point, $\Omega_+=\Omega_-=\Omega_\mathrm{EP}$ and $\gamma_+=\gamma_-=\gamma_\mathrm{EP}$, with $\Omega_\mathrm{EP}/2\pi=10.1054~\mathrm{GHz}$ and $\gamma_\mathrm{EP}/2\pi=18.7~\mathrm{MHz}$. The difference between the measured spectrum and this EP reference reflects the linewidth bifurcation beyond the branch point: the damping rate of one eigenbranch is reduced below $\gamma_\mathrm{EP}$ while the other is increased above it. Thus, $P_3$ accesses the far side of the EP within a finite stable detuning window bounded by phonon lasing. The associated branch-point topology is revealed by mapping the eigenfrequencies across the full $(P_{\rm in}, \Delta)$ parameter space.

Fig.~\ref{fig:Fig4}\textbf{a} shows the measured and analytically predicted real eigenfrequencies $\Omega_\pm$ across the $(P_{\rm in}, \Delta)$ plane. At low $P_{\rm in}$, the two sheets are separated by the bare mechanical splitting $\Delta\Omega$. As $P_{\rm in}$ increases, optically mediated coupling pulls the sheets together until they meet at the EP. Beyond this point, the continuously evolved eigenbranches reconnect onto the opposite frequency sheets. This sheet exchange is the measured real-frequency projection of the fitted complex Riemann-sheet topology characteristic of a mechanical EP.

The spectral signature of this topology is isolated in Fig.~\ref{fig:Fig4}\textbf{b} by co-varying $P_{\rm in}$ and $\Delta$ along the two degeneracy contours shown in Fig.~\ref{fig:Fig3}\textbf{d}. For $P_{\rm in}<P_{\rm EP}$, the trajectory follows the damping-degeneracy contour, $\gamma_+=\gamma_-$. Along this path, ${\rm Im}[g_1g_2\chi]$ compensates the mismatch between ${\rm Im}[\tilde{\Omega}^{\rm OM}_1]$ and ${\rm Im}[\tilde{\Omega}^{\rm OM}_2]$, balancing the damping rates and allowing the collapse of the real-frequency splitting to be resolved as $P_{\rm in}\rightarrow P_{\rm EP}$. For $P_{\rm in}>P_{\rm EP}$, the trajectory follows the frequency-degeneracy contour, $\Omega_+=\Omega_-$. Along this path, the real frequencies remain coalesced while the fitted damping rates separate, continuing the beyond-EP evolution identified in Fig.~\ref{fig:Fig3}\textbf{e} until phonon lasing appears at the highest power shown. The two contours meet at the inferred EP, shown by the purple trace, where the fitted model satisfies $\Omega_+=\Omega_-$ and $\gamma_+=\gamma_-$ simultaneously.

\subsection*{Outlook} \label{sec:disc}

\textit{Towards non-reciprocal phonon transport and state transfer.}
The exceptional point demonstrated here is an essential prerequisite for accessing topological dynamics. Quasi-adiabatic encirclement of a mechanical EP of membrane resonators has enabled mode switching and direction-dependent energy transfer~\cite{xu2016_topologicalET, Doppler2016DynamicallyEncirclingEP} that is robust to parameter fluctuations, making it well suited for implementation with nanofabricated on-chip devices. 
The high frequency of diamond OMC mechanical resonances allows cryogenic cooling to their quantum ground state. At low temperature their mechanical decoherence is also reduced by orders of magnitude \cite{ref:li2024ultracoherent, joe2024_highq_diamondomc_siv, Oh2026SpinEmbeddedDiamondOMR}, and they are less susceptible to optical heating thanks to diamond's low linear and nonlinear absorption. These properties make these devices ideal platforms for implementing topological optomechanics in the quantum regime. More broadly, expanding the multimode nature of our OMC will enable studies of synthetic gauge fields~\cite{Mathew2020SyntheticGaugeFields}, chiral phonon transport~\cite{delPino2022NonHermitianChiralPhononics}, and topologically protected phononic edge states~\cite{Slim2024BosonicKitaevChain} previously demonstrated in silicon nano-optomechanics. Advancing from EP observation to dynamic encirclement will require time-dependent modulation of laser detuning and power along a closed trajectory around the EP; the stability analysis in Fig.~\ref{fig:Fig3}\textbf{b} guides the choice of phonon-lasing-free paths. 

\textit{Driving extended resonator modes for spin-optomechanics}. Creating spin--optomechanical transducers is a key application of diamond cavity optomechanical devices~\cite{shandilya2022_diamond}. A central challenge is that the optical fields used to drive mechanical resonators can degrade the coherence of diamond spins~\cite{joe2024_highq_diamondomc_siv, ref:shandilya2024nonlinear}. It is therefore desirable to engineer mechanical resonances that generate strain away from optical hot spots, while retaining sufficient optomechanical coupling for efficient optical driving~\cite{Lodde2024StrainCouplingExcitonNanoOMR}. More generally, the mechanical mode most favorable for optomechanical excitation is not necessarily the same mode that couples most strongly to an embedded spin. The multimode diamond OMC studied here addresses this constraint by making the mechanical strain profile optically tunable.

Phase-sensitive interference between mechanical modes provides a design handle for shaping local mechanical response~\cite{ElSayed2020QNMElasticPurcellFano}. In the symmetry-broken OMC nanobeam studied here, the two mechanical modes are coherent combinations of breathing-like cavity motion and flexural-like mirror-region motion with opposite relative phase. Optically mediated hybridization can therefore tune interference between these parent components. One optically dressed eigenmode can acquire stronger optomechanical overlap with the cavity region, while the other can retain larger strain participation in the mirror region, where spin defects can be positioned with reduced exposure to the optical field. For the uncoupled mechanical modes, simulations indicate mirror-region spin--strain coupling rates of up to $g_{\rm sm,1}/2\pi=1.18~\mathrm{MHz}$ and $g_{\rm sm,2}/2\pi=1.24~\mathrm{MHz}$ to an SiV defect (Supplementary Information~\ref{supp:sims}). At the experimentally accessed point with the largest contrast in projected optomechanical coupling, $(g_{\rm eff}^{(+)},g_{\rm eff}^{(-)})/2\pi=(177,372)~\mathrm{kHz}$, projecting the corresponding eigenvector coefficients onto the simulated strain fields gives $(g_{\rm sm}^{(+)},g_{\rm sm}^{(-)})/2\pi=(1.53,0.73)~\mathrm{MHz}$. Together, these projections indicate that the optically mediated hybridization demonstrated here provides a tunable trade-off between optical drive strength and spin--strain participation in a symmetry-mixed diamond OMC. Reducing the bare mechanical splitting $\Delta\Omega$ in future devices would lower the power required to rotate the eigenvectors further towards optically and spin-strain bright limits, and would simultaneously widen the stable detuning window beyond the EP. This tunable interference connects to recent demonstrations of spin--strain control through interference between driven mechanical modes~\cite{Lee2025TuningStrainCoupling}; in our system, the same interference principle is made optically tunable through non-Hermitian hybridization of the mechanical eigenmodes.

Dynamic encirclement of the exceptional point could add a path-dependent layer to this optically tunable trade-off between $g_{\rm eff}$ and $g_{\rm sm}$. The eigenvector control demonstrated here is static: each operating point $(P_{\rm in},\Delta)$ sets a fixed partition of the optical and spin--strain projections between the two eigenbranches. Dynamically encircling $\chi_{\rm EP}$ along a closed trajectory would instead use the branch-point topology to make the evolution of mechanical excitation between branches direction dependent. In spin-embedded diamond OMCs, this could provide a route toward direction-biased phonon-mediated coupling between optical drive and spin-active strain profiles.

\section*{Acknowledgments}
We thank Thiago P.\ Mayer Alegre for helpful correspondence. We also acknowledge the authors of Ref.~\cite{Moraes2022OptimizationDiamondOMC} for making their COMSOL files publicly available, which provided the starting point of our finite-element modeling. This work was supported by the  and the Natural Science and Engineering Research Council (Discovery Grant, Research Tools and Instruments, Alliance Quantum ALLRP 586298-2023, the CanQuEST Alliance Quantum Consortia, and the ARAQNE Alliance Quantum Consortia) and the NRC-University of Alberta Nanotechnology Initiative.

\putbib[DiamondOMCs]
\end{bibunit}

\clearpage
\begin{bibunit}
\onecolumngrid
\appendix
\begin{center}
{\large\bfseries Supplementary Information}
\end{center}

\subsection{Derivation of the effective mechanical Hamiltonian}
\label{supp:heff}

The effective Hamiltonian $H_{\rm eff}$ governing the two coupled mechanical modes is derived following similar treatments in Refs.~\cite{Shkarin2014, xu2016_topologicalET, wu2023_onchipEP}; the key steps are outlined below.

The optomechanical system consists of a single optical cavity mode $\hat{a}$ dispersively coupled to two mechanical modes, $\hat{b}_1$ and $\hat{b}_2$. In a frame rotating at the laser frequency $\omega_L$, the system Hamiltonian is
\begin{equation}
\frac{\hat{H}}{\hbar} = -\Delta\, \hat{a}^\dagger \hat{a} + \sum_{j=1,2}\Omega_j \hat{b}_j^\dagger \hat{b}_j - \hat{a}^\dagger \hat{a}\sum_{j=1,2} g_j\bigl(\hat{b}_j + \hat{b}_j^\dagger\bigr) + i\sqrt{\tfrac{\kappa_{\rm ex}}{2}}\,\bigl(\alpha_{\rm in}\hat{a}^\dagger - \alpha_{\rm in}^*\hat{a}\bigr),
\label{eq:H_full}
\end{equation}
where $\Delta = \omega_L - \omega_c$ is the laser detuning from the bare cavity resonance, $\Omega_j$ and $g_j$ ($j = 1, 2$) are the bare mechanical frequencies and vacuum optomechanical coupling rates, $\kappa_{\rm ex}$ is the total fiber--cavity coupling rate, and $|\alpha_{\rm in}|^2 = P_{\rm in}/\hbar\omega_L$ is the driving photon flux.

Coupling each mode to its thermal bath and including the bath--system interaction yields the quantum Langevin equations
\begin{align}
\dot{\hat{a}} &= \bigl(i\Delta - \tfrac{\kappa}{2}\bigr)\hat{a} + i\sum_{j=1,2} g_j \hat{a}\bigl(\hat{b}_j + \hat{b}_j^\dagger\bigr) + \sqrt{\tfrac{\kappa_{\rm ex}}{2}}\,\alpha_{\rm in} + \sqrt{\kappa}\,\hat{a}_{\rm in}, \label{eq:langevin_a} \\
\dot{\hat{b}}_j &= \bigl(-i\Omega_j - \tfrac{\gamma_j}{2}\bigr)\hat{b}_j + ig_j\hat{a}^\dagger\hat{a} + \sqrt{\gamma_j}\,\hat{b}_{{\rm in},j}, \label{eq:langevin_b}
\end{align}
where $\gamma_j$ are the intrinsic mechanical dissipation rates set by the bath couplings, $\kappa = \kappa_{\rm ex} + \kappa_i$ is the total cavity decay rate, and $\hat{a}_{\rm in}$, $\hat{b}_{{\rm in},j}$ are the input noise operators associated with the optical and mechanical baths, respectively.

For a strong coherent intracavity field we write $\hat{a} = \alpha + \delta\hat{a}$ and $\hat{b}_j = \beta_j + \delta\hat{b}_j$, where $\alpha$ and $\beta_j$ are the steady-state optical and mechanical amplitudes and $\delta\hat{a}$, $\delta\hat{b}_j$ are the corresponding fluctuations. Setting the time derivatives in Eqs.~\ref{eq:langevin_a}--\ref{eq:langevin_b} to zero, dropping noise terms, and choosing the optical phase such that $\alpha = \sqrt{n_{\rm cav}}$ is real, the steady-state mechanical displacement is
\begin{equation}
    \beta_j = \frac{ig_j n_{\rm cav}}{i\Omega_j + \gamma_j/2},
\label{eq:beta_j}
\end{equation}
with intracavity photon number $n_{\rm cav} = |\alpha|^2$. The associated static radiation-pressure displacement shifts the effective cavity resonance, but for our device parameters this shift is several orders of magnitude smaller than $\kappa$ and the GHz-scale detuning range probed in the experiment; we therefore absorb it into the experimentally defined detuning $\Delta$.

Linearizing Eqs.~\ref{eq:langevin_a}--\ref{eq:langevin_b} about this steady state and retaining only terms first-order in the fluctuations gives the linearized Langevin equations
\begin{align}
\delta\dot{\hat{a}} &= \bigl(i\Delta - \tfrac{\kappa}{2}\bigr)\delta\hat{a} + i\sum_{j=1,2} G_j\bigl(\delta\hat{b}_j + \delta\hat{b}_j^\dagger\bigr) + \sqrt{\kappa}\,\hat{a}_{\rm in}, \label{eq:lin_a} \\
\delta\dot{\hat{b}}_j &= \bigl(-i\Omega_j - \tfrac{\gamma_j}{2}\bigr)\delta\hat{b}_j + iG_j\bigl(\delta\hat{a} + \delta\hat{a}^\dagger\bigr) + \sqrt{\gamma_j}\,\hat{b}_{{\rm in},j}, \label{eq:lin_b}
\end{align}
where $G_j = g_j\sqrt{n_{\rm cav}}$ is the photon-enhanced optomechanical coupling.

Solving Eqs.~\ref{eq:lin_a}--\ref{eq:lin_b} in the frequency domain for $\delta\hat{a}(\omega)$ in terms of $\delta\hat{b}_j(\omega)$ adiabatically eliminates the cavity field. Substituting back into Eq.~\ref{eq:lin_b} and evaluating the optical response near the mechanical resonance frequencies (such that rapidly rotating terms can be dropped) yields cavity-mediated couplings governed by a single complex susceptibility,
\begin{equation}
\chi(\Delta) = n_{\rm cav}\left[\bigl(\omega + \Delta + i\tfrac{\kappa}{2}\bigr)^{-1} - \bigl(\omega - \Delta + i\tfrac{\kappa}{2}\bigr)^{-1}\right],
\label{eq:chi}
\end{equation}
with $\omega = (\Omega_1 + \Omega_2)/2$. The mechanical fluctuations then obey
\begin{equation}
i\frac{d}{dt}\delta\mathbf{b} = H_{\rm eff}\,\delta\mathbf{b}, \qquad \delta\mathbf{b} = (\delta\hat{b}_1, \delta\hat{b}_2)^T,
\label{eq:b_eom}
\end{equation}
with
\begin{equation}
H_{\rm eff} = \begin{pmatrix} \Omega_1 - i\gamma_1/2 + g_1^2\chi(\Delta) & g_1 g_2\chi(\Delta) \\ g_1 g_2\chi(\Delta) & \Omega_2 - i\gamma_2/2 + g_2^2\chi(\Delta) \end{pmatrix}.
\label{eq:Heff_S}
\end{equation}
This is the effective Hamiltonian quoted in Eq.~2 of the main text. For compactness we hereafter write $\mathbf{b} = (\hat{b}_1, \hat{b}_2)^T$; the operators $\hat{b}_j$ should be understood as fluctuations about the static mechanical displacement.

Denoting the optically shifted single-mode complex frequencies as
\begin{equation}
\tilde{\Omega}_j^{\rm OM} = \Omega_j - i\frac{\gamma_j}{2} + g_j^2\chi(\Delta),
\label{eq:Omega_OM}
\end{equation}
the eigenfrequencies of Eq.~\ref{eq:Heff} are
\begin{equation}
\tilde{\Omega}_\pm = \tfrac{1}{2}\bigl(\tilde{\Omega}_1^{\rm OM} + \tilde{\Omega}_2^{\rm OM}\bigr) \pm \tfrac{1}{2}S,
\label{eq:Omega_pm}
\end{equation}
with complex splitting
\begin{equation}
S = \sqrt{\bigl(\tilde{\Omega}_1^{\rm OM} - \tilde{\Omega}_2^{\rm OM}\bigr)^2 + 4\bigl(g_1 g_2\chi\bigr)^2}.
\label{eq:S}
\end{equation}
An exceptional point corresponds to $S = 0$, at which both the eigenvalues and eigenvectors of $H_{\rm eff}$ coalesce.

\subsection{Projected optomechanical coupling}
\label{supp:geff}

The right eigenvectors associated with $\tilde{\Omega}_\pm$ satisfy $H_{\rm eff}\mathbf{u}^{(\pm)} = \tilde{\Omega}_\pm \mathbf{u}^{(\pm)}$, with normalization $|u_1^{(\pm)}|^2 + |u_2^{(\pm)}|^2 = 1$. Their components are
\begin{equation}
    u_1^{(\pm)} = \frac{1}{\sqrt{1 + |r_\pm|^2}}, \qquad u_2^{(\pm)} = \frac{r_\pm}{\sqrt{1 + |r_\pm|^2}},
    \label{eq:u_components}
\end{equation}
with
\begin{equation}
    r_\pm = \frac{\tilde{\Omega}_\pm - \tilde{\Omega}_1^{\rm OM}}{g_1 g_2\chi}.
\end{equation}
The overall complex phase of $\mathbf{u}^{(\pm)}$ is arbitrary and does not affect the projected couplings defined below.

The radiation-pressure interaction in Eq.~\eqref{eq:H_full} couples the cavity field to the collective mechanical coordinate
\begin{equation}
\sum_{j=1,2} g_j\bigl(\hat{b}_j + \hat{b}_j^\dagger\bigr) = (\mathbf{g}\cdot\mathbf{b}) + (\mathbf{g}\cdot\mathbf{b})^\dagger,
\end{equation}
with optomechanical coupling vector
\begin{equation}
    \mathbf{g} = (g_1, g_2)^T.
    \label{eq:g_vector}
\end{equation}
The effective coupling of each hybridized mechanical eigenmode to the optical field is therefore set by the projection of its eigenvector onto $\mathbf{g}$:
\begin{equation}
    g_{\rm eff}^{(\pm)} = |\mathbf{g}\cdot\mathbf{u}^{(\pm)}| = |g_1 u_1^{(\pm)} + g_2 u_2^{(\pm)}|.
    \label{eq:g_eff_S}
\end{equation}

Maximizing and minimizing $|\mathbf{g}\cdot\mathbf{u}|$ over normalized mechanical vectors defines the optically bright and dark directions, respectively. For real $g_1, g_2$, the bright eigenvector is
\begin{equation}
    \mathbf{u}_B = \frac{1}{\sqrt{g_1^2 + g_2^2}}\,(g_1, g_2)^T, \qquad g_{\rm eff,B} = \sqrt{g_1^2 + g_2^2},
\end{equation}
and the orthogonal dark direction is
\begin{equation}
    \mathbf{u}_D = \frac{1}{\sqrt{g_1^2 + g_2^2}}\,(g_2, -g_1)^T, \qquad g_{\rm eff,D} = 0.
\end{equation}
In the balanced limit $g_1 \simeq g_2$ these reduce to the symmetric and antisymmetric combinations $\mathbf{u}_B \simeq (1, 1)^T/\sqrt{2}$ and $\mathbf{u}_D \simeq (1, -1)^T/\sqrt{2}$.

The bright and dark directions are determined by the coupling vector $\mathbf{g}$ alone, while the hybridized eigenvectors $\mathbf{u}^{(\pm)}$ depend on $\chi(\Delta)$. As $\chi$ is tuned in our experiment, $\mathbf{u}^{(\pm)}$ rotate within the $(\mathbf{u}_B, \mathbf{u}_D)$ basis, redistributing $g_{\rm eff}$ between the two branches. Because $H_{\rm eff}$ is non-Hermitian, $\mathbf{u}^{(+)}$ and $\mathbf{u}^{(-)}$ are not in general orthogonal, and this redistribution is not bounded by a simple sum rule.

At an EP, the eigenvector components in Eq.~\ref{eq:u_components} satisfy
\begin{equation}
    \frac{u_{2, \rm EP}}{u_{1,\rm EP}} = \pm i,
\end{equation}
with the sign determined by which EP branch (red- or blue-detuned) is reached. Substituting into Eq.~\eqref{eq:g_eff_S} yields the projected coupling at the EP,
\begin{equation}
    g_{\rm eff,EP} = \frac{1}{\sqrt{2}}|g_1 \pm i g_2| = \sqrt{\frac{g_1^2 + g_2^2}{2}},
\end{equation}
where the two coalescing branches share the same effective coupling, equal to the root-mean-square of the bare optomechanical coupling rates.

\subsection{Conditions for stable Exception Points} \label{supp:stability}
\begin{figure}[h]
	\includegraphics[width=0.6\linewidth]{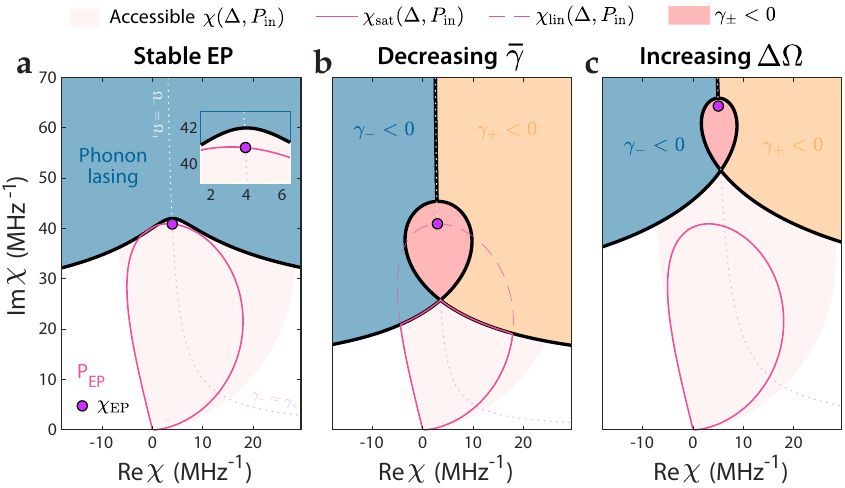}
		\caption{
            \small \textbf{Stability of the exceptional-point trajectory.} ({\bf a}) Stability map of the complex cavity susceptibility $\chi={\rm Re}[\chi]+i\,{\rm Im}[\chi]$ for the same experimentally extracted parameters in the main text, with $|\Delta\Omega|/2\pi=6.9~\mathrm{MHz}$ and $\bar{\gamma} = 8.8~\mathrm{MHz}$ . Shaded regions indicate the unstable regions where ${\rm Im}[\tilde{\Omega}_{\pm}]>0$. The white and purple dotted lines show where ${\rm Re}[\tilde{\Omega}_{+}]={\rm Re}[\tilde{\Omega}_{-}]$ and ${\rm Im}[\tilde{\Omega}_{+}]={\rm Im}[\tilde{\Omega}_{-}]$, respectively. Their intersection defines the EP. The pink shaded lobe denotes the experimentally accessible values of $\chi(P_{\rm in}, \Delta)$ for the measured cavity coupling ($\kappa_{\mathrm{ex}},\kappa$) and available laser power, showing that the EP lies within both the stable and accessible regime. ({\bf b}) Same as ({a}) with intrinsic mechanical dissipation rates $\bar{\gamma} \to 0.5\bar{\gamma}$ reduced by half, which shifts the lasing threshold down in the $\chi$-plane below $\chi_{\rm EP}$. In this regime one mode lases before the two modes coalesce, forcing $\chi(\Delta)$ to saturate along the instability boundary where ${\rm Im}[\tilde{\Omega}_{\pm}]=0$. Similarly, increasing the mechanical-mode frequency splitting  ($|\Delta\Omega| \to 11~\mathrm{MHz}$) ({\bf c}) shifts the EP up into the self-oscillatory region.
            }
            \label{fig:contour}
\end{figure}
When optomechanical backaction amplification overcomes the intrinsic mechanical damping of a hybridized eigenmode, that mode undergoes self-oscillation, developing a coherent mechanical amplitude
\begin{equation}
    \delta\mathbf{b}_\lambda(t) = A_\lambda\,\mathbf{u}^{(\lambda)}e^{-i\omega_\lambda t},
\end{equation}
where $A_\lambda$ is the oscillation amplitude, $\mathbf{u}^{(\lambda)}$ is the hybridized eigenvector of the self-oscillating branch, and $\lambda \in \{+,-\}$. In the linearized model, $A_\lambda$ grows without bound. In practice, the large mechanical oscillation strongly modulates the cavity resonance at $\omega_\lambda$, redistributing the intracavity field into higher-order sidebands and thereby reducing the net optical gain available to drive the motion. The oscillation amplitude therefore saturates at a steady-state limit cycle for which the effective damping of the self-oscillating branch vanishes, $\gamma_\lambda = 0$. In this regime the linearized model of Section~\ref{supp:heff} breaks down, and the cavity susceptibility is replaced by a saturated value $\chi \to \chi_{\rm sat}$, which clamps the trajectory of $\chi(P_{\rm in}, \Delta)$ to the phonon-lasing threshold contour $\gamma_\pm = 0$ as $\Delta$ is swept. The effective Hamiltonian retains the form of Eq.~\ref{eq:Heff_S}, but with $\chi$ replaced by $\chi_{\rm sat}$. Following Ref.~\cite{wu2023_onchipEP}, this saturated susceptibility is
%
\begin{equation}
    \chi_{\rm sat} =  \frac{\kappa_{\rm ex}|\alpha_{\rm in}|^2}{z_\lambda\,\omega_\lambda} \sum_{n=-\infty}^{\infty} \frac{J_n(z_\lambda)\,J_{n+1}(z_\lambda)}{\left[i(n\omega_\lambda - \Delta) + \frac{\kappa}{2}\right]\!\left[-i((n+1)\omega_\lambda - \Delta) + \frac{\kappa}{2}\right]},
\end{equation}

where $J_n$ is the $n$-th order Bessel function of the first kind, $\omega_\lambda = {\rm Re}[\tilde{\Omega}_\lambda]$ is the self-oscillation frequency, and
\begin{equation}
    z_\lambda = \frac{2|A_\lambda|\,g_{\rm eff}^{(\lambda)}}{\omega_\lambda}
\end{equation}
is the modulation index, determined by the oscillation amplitude and the projected optomechanical coupling $g_{\rm eff}^{(\lambda)} = |\mathbf{g}^{T}\mathbf{u}^{(\lambda)}|$ of the self-oscillating branch. In fitting the self-oscillating portions of the data, $z_\lambda$ is determined self-consistently at each detuning by enforcing $\gamma_\lambda(\chi_{\rm sat}) = 0$: once the linearized model predicts $\gamma_\lambda < 0$, $z_\lambda$ is increased until the saturated susceptibility restores zero net damping. The susceptibility trajectory $\chi(P_{\rm in}, \Delta)$ therefore follows the phonon-lasing threshold contour as $\Delta$ is swept wherever the linear trajectory would enter the unstable region. This gain clamping can prevent access to the EP if $\chi_{\rm EP}$ lies above the instability threshold.

Fig.~\ref{fig:contour} illustrates how the device parameters determine whether the blue-detuned EP can be accessed before the onset of self-oscillation. As $P_{\rm in}$ increases, the susceptibility trajectory expands in the complex $\chi$-plane toward the instability boundary; once reached, $\chi$ is clamped to the threshold contour at those detunings where the linear trajectory would cross into the unstable region. The EP occurs when the complex splitting $S$ vanishes at the condition:
\begin{equation}
    \chi(P_{\rm in}, \Delta) = \chi_{\rm EP} \equiv  \frac{\Delta\Omega - \tfrac{i}{2}\Delta\gamma}{(g_2^2 - g_1^2) \pm i\,2 g_1 g_2}
\end{equation}
where $\Delta\Omega = \Omega_1 - \Omega_2$ and $\Delta\gamma = \gamma_1 - \gamma_2$. The sign of ${\rm Im}[\chi_{\rm EP}]$ determines whether the EP occurs for a red-detuned (${\rm Im}[\chi_{\rm EP}] < 0$) or blue-detuned (${\rm Im}[\chi_{\rm EP}]>0$) drive laser. If the relevant blue-detuned solution $\chi_{\rm EP}$ lies in the unstable region, the EP is experimentally inaccessible. For the extracted parameters of the present device, Fig.~\ref{fig:contour}\textbf{a} shows that the blue-detuned EP lies within both the stable region and the experimentally accessible range of $\chi(P_{\rm in}, \Delta)$. Figures~\ref{fig:contour}\textbf{b,c} show how this accessibility depends sensitively on the intrinsic mechanical damping and bare frequency splitting: reducing $\bar{\gamma} = (\gamma_1 + \gamma_2)/2$ lowers the phonon-lasing threshold in ${\rm Im}[\chi]$, while increasing $\Delta\Omega$ shifts $\chi_{\rm EP}$ to larger $|\chi|$, in both cases causing one branch to self-oscillate before the eigenvalues coalesce. The condition for the blue-detuned EP to remain stable is
\begin{equation}
    \bar{\gamma}(g_1^2 + g_2^2) + \frac{1}{2}(g_2^2 - g_1^2)\Delta\gamma > \left|2 g_1 g_2\,\Delta\Omega\right|,
    \label{eq:EPstability}
\end{equation}
which expresses the competition between the intrinsic mechanical damping, which stabilizes the blue-detuned system against phonon lasing, and the optically mediated coupling strength required to reach the EP. In the balanced limit $g_1 \simeq g_2$ and $\Delta\gamma \simeq 0$, the condition reduces to $\bar{\gamma} > |\Delta\Omega|$, showing that stable EP access requires the bare mode splitting to be small compared with the mechanical damping.


\subsection{Diamond optomechanical crystal design and simulation}  \label{supp:sims}
\begin{figure*}[t]
	\includegraphics[width=1\linewidth]{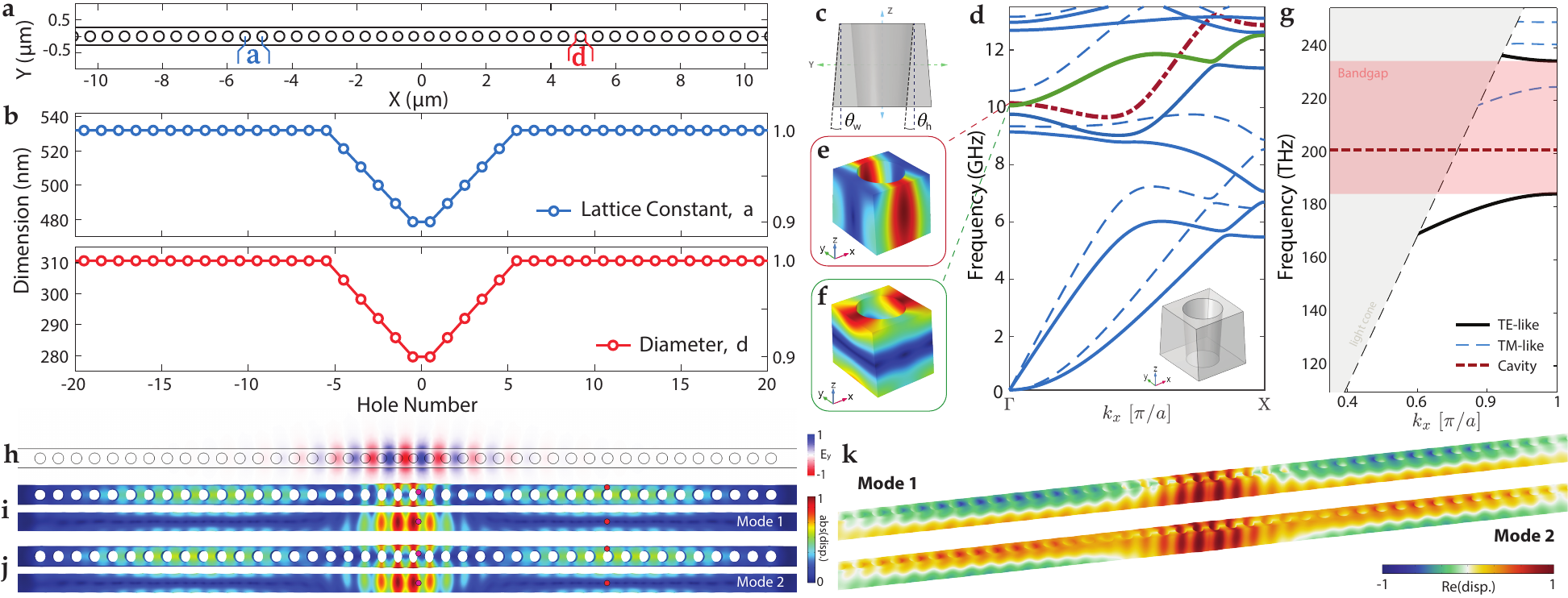}
		\caption{
            \small \textbf{Diamond optomechanical crystal design.} (\textbf{a}) Nanobeam geometry in the $xy$-plane at $z = 0$, consisting of a periodic array of holes with lattice constant $a$ and hole diameter $d$. The beam thickness and width are 530.2 nm and 557.0 nm, respectively. (\textbf{b}) The mirror-lattice dimensions ($a = 531.9$ nm, $d = 319.1$ nm) are linearly tapered to a defect lattice at the beam center ($a = 452.1$ nm, $d = 271.3$ nm), forming simultaneous optical and mechanical cavities. ({\bf c}) Cross-section of the OMC unit cell, with sidewall angles $\theta_{\rm w} = 3^{\circ}$ and $\theta_{\rm h} = 2^{\circ}$, which break vertical reflection symmetry. (\textbf{d}) Mechanical band structures of the mirror (solid) and defect (dashed) lattices. The mirror-lattice and cavity bands of interest are highlighted in green and red, respectively. At the $\Gamma$-point, these bands correspond to a cavity breathing mode (\textbf{e})and a mirror-lattice out-of-plane flexural mode (\textbf{f}). (\textbf{g}) Optical band structure. (\textbf{h}) Normalized $E_{y}$ profile of the optical cavity mode, simulated at $\lambda_0 = 1563$ nm with quality factor $Q_{\rm opt} = 1.5\times10^5$. (\textbf{i, j}) Top and side views of the normalized displacement profiles of the two symmetry-mixed mechanical supermodes, simulated at $\Omega_{1}/2\pi = 10.293~\mathrm{GHz}$ and $\Omega_{2}/2\pi = 10.300~\mathrm{GHz}$. (\textbf{k}) Real displacement profiles of the same mechanical modes, showing out-of-phase and in-phase combinations of the cavity breathing and mirror-region flexural motions.
            }
            \label{fig:sims}
\end{figure*}
The diamond optomechanical crystal studied consists of a suspended nanobeam perforated by a one-dimensional lattice of circular holes. The hole sidewalls have an angle of $\theta_h = 2^\circ$ while the beam's outer walls have an angle of $\theta_w = 3^\circ$. The lattice constant and hole diameter of nominal unit cell in the mirror region are tapered toward the beam center to form a simultaneous optical and mechanical defect cavity, as shown in Fig.~\ref{fig:sims}. This taper creates a localized optical cavity mode within a photonic bandgap and simultaneously pulls selected mechanical resonances from the mirror lattice into the cavity region. The mechanical and optical properties were simulated using finite-element analysis in COMSOL. Band structures were obtained using Floquet periodic boundary conditions along the nanobeam axis $x$. Fig.~\ref{fig:sims}\textbf{g} shows the nominal optical band structure, from which the defect is designed to shift the transverse electric (TE-like) valence band at $k_x=1$ up into the bandgap. The resulting optical field profile is shown in Fig.~\ref{fig:sims}\textbf{h}. 

Fig.~\ref{fig:sims}\textbf{d} shows the simulated mechanical bands of the mirror lattice together with the bands of the defect lattice. The two highlighted branches from the defect and nominal lattices provide the two underlying modes of interest. In the absence of a sidewall angle, these two mechanical modes belong to distinct symmetry sectors and do not interact. The first is a cavity breathing mode, which is predominantly dilatational and produces strong volumetric strain in the cavity region, exhibiting strong optomechanical coupling~\cite{chan_optimized_2012}. Its displacement components, which are primarily dominated by $u_y$, transform as:
\begin{equation}
\begin{aligned}
        u_x &: (\text{odd in }x, \text{ even in }y, \text{ even in }z) \\
        u_y &: (\text{even in }x, \text{ odd in }y, \text{ even in }z) \\
        u_z &: (\text{even in }x, \text{ even in }y, \text{ odd in }z)
\end{aligned}
\end{equation}
The second is a nearly degenerate mirror-lattice mode that is predominantly out-of-plane flexural. Its dominant displacement is $u_z$, with  
\begin{equation}
\begin{aligned}
        u_x &: (\text{even in }x, \text{ even in }y, \text{ odd in }z) \\
        u_y &: (\text{odd in }x, \text{ odd in }y, \text{ odd in }z) \\
        u_z &: (\text{odd in }x, \text{ even in }y, \text{ even in }z)
\end{aligned}
\end{equation}
Accordingly, the flexural mode is odd along the beam axis and bending-like, whereas the breathing mode is localized and compressional. Introducing sidewall angles in the structure breaks the vertical ($z$) reflection symmetry about the $z=0$ plane. This asymmetry activates coupling between the otherwise symmetry-protected cavity breathing mode and the mirror-region flexural mode: their overlap no longer cancels by reflection symmetry and the two modes mix. This produces the two co-localized mechanical supermodes shown in Fig.~\ref{fig:sims}\textbf{i--k}, separated by $|\Delta\Omega|/2\pi \approx 7~\mathrm{MHz}$. One supermode is an in-phase combination, for which the cavity breathing displacement and the mirror-region flexural displacement have the same sign, while the other is an out-of-phase combination, for which these contributions have opposite sign. 

Physically, this sidewall angle acts as a controlled perturbation on an otherwise nearly symmetric structure. This regime is distinct from strongly asymmetric diamond nanobeams, where large cross-sectional asymmetry can shift the underlying breathing-like and flexural-like modes far apart in frequency and thereby suppress strong mixing. In the present device, the sidewall angle, along with the lattice and taper design, keeps these mode families nearly degenerate in the full OMC structure while still inducing appreciable mixing. As a result, both modes inherit the breathing character in the cavity region required for strong optomechanical coupling to the optical mode; without this mixing, one mode would remain primarily mirror-localized and interact only weakly with the optical field. Although these mechanical modes are not protected by a complete phononic bandgap, they remain instead quasi-localized by the patterned mirror region, which produces strong reflection and poor mode matching to the propagating modes of the adjacent uniform beam.

The simulated optomechanical coupling rates of the two mixed-symmetry modes are  $g_{1}/2\pi = 262~\mathrm{kHz}$ and $g_{2}/2\pi = 259~\mathrm{kHz}$. For mode 1, the moving-boundary and photoelastic contributions are $g_{1,\rm MB}/2\pi = 87~\mathrm{kHz}$ and $g_{1,\rm PE}/2\pi = 175~\mathrm{kHz}$, respectively; for mode 2, they are $g_{2,\rm MB}/2\pi = 86~\mathrm{kHz}$ and $g_{2,\rm PE}/2\pi = 173~\mathrm{kHz}$. 

Owing to the modest simulated mechanical quality factors, $Q_{\rm m,1} = 9.1\times10^3$ and $Q_{\rm m,2} = 8.4\times10^3$, we compute the optomechanical coupling rates using a quasinormal-mode (QNM) normalization, with mechanical effective mass $m_{\rm eff} \equiv \rho \Re(V_{\rm eff}) $, where $V_{\rm eff}$ is the complex QNM-normalized effective mode volume; the full normalization expression is given in Refs.~\cite{ElSayed2020QNMElasticPurcellFano}. We note, however, that for modes with this degree of loss, the standard normal-mode normalization remains a good approximation, yielding $g_{1, \rm NM}/2\pi = 261~\mathrm{kHz}$, $g_{2, \rm NM}/2\pi = 257~\mathrm{kHz}$)~\cite{eichenfield_modeling_2009}. The finite-element models used here were adapted from COMSOL files made available by the authors of Ref.~\cite{Moraes2022OptimizationDiamondOMC}. 

The spin--mechanical coupling rates ($g_{\rm sm}$) to a silicon-vacancy (SiV$^{-}$) center were estimated in COMSOL by projecting the simulated mechanical strain tensor onto the SiV$^{-}$ transverse strain component, following the methodology and publicly available simulation files of Raniwala \textit{et al.}~\cite{Raniwala2025}. The fabricated OMC nanobeam is aligned along the diamond $[100]$ crystallographic axis, and we evaluate the coupling for a representative $[111]$-oriented SiV$^{-}$ defect. The quoted values correspond to the zero-point-normalized strain-orbital coupling and should be interpreted as an upper-bound estimate for the spin--phonon coupling, which can be approached under an appropriate applied magnetic field. The actual spin-transition coupling depends on the SiV orientation, local static strain, and applied magnetic-field configuration, as described in more complete spin--phonon models~\cite{Joe2026PurcellEnhancedSpinPhonon}. Evaluating the coupling at the mode maximum in the central cavity region, indicated by the magenta dots in Fig.~\ref{fig:sims}\textbf{i,j}, gives maximum coupling rates of $g_{\rm sm,1}/2\pi = 4.54~\mathrm{MHz}$ and $g_{\rm sm,2}/2\pi = 4.40~\mathrm{MHz}$. To reduce optically induced spin decoherence and heating, the spin can instead be positioned in the mirror region, away from regions of high optical intensity. Because the symmetry-mixed mechanical modes extend into this region, the maximum coupling rates at the locations indicated by the red dots in Fig.~\ref{fig:sims}\textbf{i,j} remain $g_{\rm sm,1}/2\pi = 1.18~\mathrm{MHz}$ and $g_{\rm sm,2}/2\pi = 1.24~\mathrm{MHz}$. At the same mirror-region location, a defect implanted 60~nm below the nearest surface retains coupling rates of $g_{\rm sm,1}/2\pi = 732~\mathrm{kHz}$ and $g_{\rm sm,2}/2\pi = 808~\mathrm{kHz}$.

\subsection{Nanofabrication and sidewall angle} \label{supp:fab}

The fabrication of the diamond devices follows a quasi-isotropic inductively coupled plasma reactive ion etching (ICP-RIE) process described in detail elsewhere~\cite{Khanaliloo2015NB,mitchell_realizing_2019, zohari2022optomechanical}. The sidewall angle is inherently introduced during the two sequential anisotropic etch steps---the \ce{Si3N4} hard mask etch (\ce{C4F8}/\ce{SF6}) and the directional diamond etch (\ce{O2})---and their contributions compound to break the vertical symmetry of the final structure.

In ICP-RIE, anisotropy arises from the directional acceleration of ions across the plasma sheath toward the substrate. In practice, ions arrive at the surface with a small spread of angles at the surface normal, as the directed velocity gained across the sheath is always finite relative to the transverse thermal velocity carried from the bulk plasma. This produces a small lateral etch component that tapers the sidewalls. The \ce{Si3N4} mask itself develops a tapered profile for the same reason, and this geometry is then transferred and amplified into the diamond during the subsequent \ce{O2} etch, compounding the net sidewall angle.

The angle is controlled primarily through the RF bias voltage applied during the two anisotropic etch steps, which sets the directed ion velocity and thus the angular spread of the ion flux. Chamber pressure provides an additional handle, with higher pressures broadening the effective ion angular distribution and increasing the sidewall angle. In the \ce{Si3N4} etch, the \ce{C4F8}/\ce{SF6} gas ratio offers further independent control through its effect on the sidewall passivation balance. Critically, because the two anisotropic steps contribute independently and additively to the net angle, their parameters can be tuned separately to provide fabrication-controlled asymmetry. This symmetry-breaking is distinct from prior work employing angled Faraday cage etching to produce triangular cross-section diamond devices~\cite{burek2016_diamondomc}.


\subsection{Measurement Setup}
\begin{figure}[h]
    \centering
\includegraphics[width=0.5\linewidth]{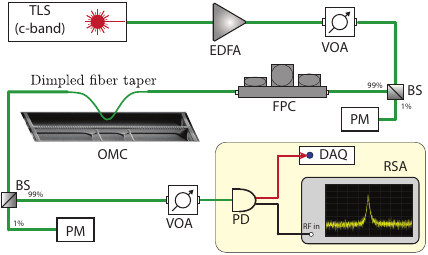}
    \caption{ \small \textbf{Experimental setup.} Schematic of the ambient measurement setup to characterize the diamond optomechanical crystal. Light from a tunable C-band laser (TLS) is amplified by an erbium-doped fiber amplifier (EDFA). The input optical power is controlled using a variable optical attenuator (VOA). Beam splitters (BS) divert part of the light to power meters (PM) for monitoring. A fiber polarization controller (FPC) optimizes the coupling of a dimpled fiber taper to the diamond OMC cavity. A second VOA attenuated the transmitted signal before it is detected on a high-speed photoreceiver (PD), whose output is fed to a data-acquisition unit (DAQ) and a real-time spectrum analyzer (RSA) for simultaneous optical and mechanical signal detection, respectively.}
    \label{fig:setup_schematic}
\end{figure}

The experimental setup used to characterize the diamond OMC is shown in Fig.~\ref{fig:setup_schematic}. Measurements were performed under ambient conditions using a dimpled fiber taper waveguide to evanescently couple light into and out of the optical cavity. Light from a widely tunable continuous-wave laser (Santec TSL-710, 1480--1640 nm) was amplified by an erbium-doped fiber amplifier (EDFA, Pritel LNHPFA-30, pre-amplifier only) and attenuated using variable optical attenuators (VOA, EXFO FVA-3100). A 99:1 fiber beam splitter (Thorlabs TW1550R1A1) directed 1\% of the input light to a power meter (Thorlabs PM400) to monitor the power injected into the fiber taper. The dimpled region of the fiber taper was positioned onto the support structures patterned adjacent to the nanobeam using stepper-motor translation stages (Suruga Seiki XXC06020-G). A fiber polarization controller was adjusted to maximize coupling to the TE-like optical cavity mode. The optical power was varied by holding the EDFA operating point fixed and stepping the attenuation of the downstream VOA. This procedure allowed the input power to be changed without modifying the laser wavelength scan or the amplifier response between measurements. 

The transmitted light was collected from the output of the fiber taper and sent through a second 99:1 beamsplitter. The 1\% port was monitored using a power meter, while the remaining signal was attenuated and sent to a high-speed photodetector (PD, Thorlabs RXM25AA). The DC photodetector signal was recorded using a data-acquisition unit (DAQ) to measure the optical transmission as the laser wavelength was swept, while the radio-frequency (RF) photocurrent was simultaneously recorded as time-domain IQ data using a real-time spectrum analyzer (RSA, Keysight N9020B MXA). The laser scan provided a trigger signal through an SMA connection to the RSA, which initiated the IQ acquisition at the start of each wavelength sweep. Each IQ trace was then divided into time windows and Fourier transformed to obtain the mechanical power spectral density  (PSD) as a function of laser detuning. 

Because the fiber-taper transmission was measured end-to-end, the optical power $P_{\rm in}$ at the cavity coupling region was estimated by assuming that the taper loss is approximately symmetric about the dimple. We therefore define
\begin{equation}
    P_{\rm in} = \sqrt{\eta_{\rm fiber}} P_{\rm L, in},
\end{equation}
where $P_{\rm L, in}$ is the power measured at the input of the fiber taper and $\eta_{\rm fiber}$ is the measured total fiber-taper efficiency. For the fiber taper used in these measurements, while mounted on the support arms, $\eta_{\rm fiber} = 0.056$ around 1550 nm. Unless otherwise stated, the input powers quoted throughout this work refer to this estimated power at the cavity coupling region.


\subsection{Fitting experimental measurements} \label{Supp:fitting}
The normalized optical transmission spectrum of the fiber taper waveguide evanescently coupled to the OMC nanobeam cavity can be described by~\cite{Mitchell2019CoherentCavityOptomechanics,Itoi2025PhotorefractiveTuningDiamondCavity}:
\begin{equation}
    T(\Delta) = \left|e^{i\phi}-\frac{\kappa_{\rm ex}/2}{\kappa/2 - i\Delta} \right|^2,
\end{equation}
where  $\phi$ accounts for Fano interference effects. The total cavity decay rate $\kappa = \kappa_{\rm i} + \kappa_{\rm ex}$ consists of intrinsic and extrinsic loss channels. The factor of 1/2 arises from the fact that cavity field couples to the forward and backward propagating fiber modes, each at a rate $\kappa_{\rm ex}$/2. At high laser powers, optical gradient forces can cause the fiber to move closer towards the cavity, changing $\kappa_{\rm ex}$. As such, when fitting the mechanical eigenfrequency shifts, $\kappa_{\rm ex}$ is extracted for each laser power, while the intrinsic decay $\kappa_{\rm i} = 8.80~\mathrm{GHz}$ is assumed to stay constant. We find a steady shift $\kappa_{\rm ex} = 6.94~\mathrm{GHz}$ at low powers, up to $\kappa_{\rm ex} = 7.03~\mathrm{GHz}$ ($P_{\rm in} = 12.1~\mathrm{mW}$) and $\kappa_{\rm ex} = 7.27~\mathrm{GHz}$ ($P_{\rm in} = 18.6~\mathrm{mW}$) at high powers. 

The intracavity photon number $n_{\rm cav}$ is given by
\begin{equation}
    n_{\rm cav} (P_{\rm in}, \Delta) = \frac{P_{\rm in}}{\hbar\omega_L}\frac{\kappa_{\rm ex}/2}{(\kappa/2)^2 + \Delta^2}.
\end{equation}
To account for thermo-optic shifts due to optical absorption, we substitute $\Delta \to \Delta_{\rm eff} = \Delta_0 - c_{T} n_{\rm cav}$, where $\Delta_0 = \omega_L - \omega_c$ is the cold-cavity detuning and $c_{T}/2\pi = 5.17~\mathrm{kHz}$ is the thermo-optic coefficient. All fitting and theoretical calculations use $\Delta_{\rm eff}$.  

For presentation, however, the detuning axis shown in figures is a statically shifted cold-cavity detuning. For each input power, we define
\begin{equation}
    \Delta \equiv \Delta_{\rm plot} = \Delta_0 - \Delta_{0,\rm ref}(P_{\rm in}),
\end{equation}
where $\Delta_{0,\rm ref}(P_{\rm in})$ is the value of $\Delta_0$ for which the fitted thermo-optic model gives $\Delta_{\rm eff} = 0$ at a given power. Thus, the plotted detuning is defined such that $\Delta = 0$ at the power-dependent thermally shifted cavity resonance, while the model itself is evaluated using the self-consistent detuning $\Delta_{\rm eff}$. This convention allows direct comparison of the fitted theory with the raw measured spectra without remapping the experimental scan axis to remove the thermo-optic distortion, while making the power-dependent mode evolution easier to compare visually.

The mechanical information of the two-mode system is transduced into an electrical voltage signal $V(t)$ via the scattered light collected at the photodetector. Following the multimode optomechanical model described in Wu et al.~\cite{wu2023_onchipEP}, the single-sided power spectral density (PSD) of the voltage signal in the frequency domain is given by:
%
\begin{equation}
\overline{S}_{pp}(\omega)
\approx \frac{H(\Delta,\kappa,\kappa_{ex})}{|\tilde{\Omega}_{+}-\omega|^{2}|\tilde{\Omega}_{-}-\omega|^{2}}\biggl[
g_{1}^{2}\gamma_{1}\bigl((\Omega_{2}-\omega)^{2}+(\gamma_{2}/2)^{2}\bigr)
+ g_{2}^{2}\gamma_{2}\bigl((\Omega_{1}-\omega)^{2}+(\gamma_{1}/2)^{2}\bigr)
\biggr].
\label{eq:S_PSD}
\end{equation}
The transduction function $H(\Delta,\kappa,\kappa_{\rm ex})$ accounts for the optical cavity's conversion of mechanical motion to the measured voltage~\cite{Gorodetsky2010VacuumG0Calibration}:
%
\begin{equation} 
    H(\Delta,\kappa,\kappa_{\rm ex}) =
     \frac{2n_{b}(\kappa_{\rm ex}GP_{\rm out})^{2}\Delta^{2}((\kappa-\kappa_{ex}/2)^{2}+w^{2})/R_{L}}{\left((\Delta^{2}+(\kappa/2)^{2} \right)^{2}((\Delta+w)^{2}+(\kappa/2)^{2})((\Delta-w)^{2}+(\kappa/2)^{2})},
\end{equation}
where $P_{\rm out}$ is the power of the transmitted light, $G$ is the conversion gain of the photodetector, $R_{L}$ is the load resistance of the spectrum analyzer, and $n_b$ is the thermal phonon occupancy. Note that we observe a slight Fano effect in the measured optical transmission, which is not accounted for in $H(\Delta,\kappa,\kappa_{\rm ex})$. As such, the detuning-dependent PSD amplitude in simulated spectrograms appears slightly shifted in frequency, despite the mode eigenfrequencies at each detuning matching with experiment. To heuristically correct for this, we apply a detuning shift of $\delta\Delta_{\rm fano} = 1.5~\mathrm{GHz}$ to the transduction function only: $H( \Delta - \delta\Delta_{\rm Fano},\kappa,\kappa_{\rm ex})$.

\subsection{Mechanical thermal effects}
In addition to the thermo-optic effect, we also observe a thermal shift in the intrinsic mechanical frequencies due to laser-induced absorption. In our analytical fitting of the power-dependent mechanical frequency trajectories, we approximate the heating induced shift to the intrinsic frequencies $\Omega_{j}$ (for $j = 1,2$) of the two mechanical modes as:
\begin{equation}
    \Omega_j(n_{\rm cav}) = \Omega_{j,0} - c_{j} n_{\rm cav}
\end{equation}
where $c_j$ are the effective thermal frequency shift coefficients for each mode. While the two mechanical modes exhibit nearly identical spatial energy distributions, their respective thermal coefficients ($c_1/2\pi = 0.151$ Hz and $c_2/2\pi = 0.044$ Hz) were found to be distinct due to the different thermal response of the underlying bare modes---the cavity-localized breathing mode and mirror-localized flexural mode---before symmetry-breaking-induced mode mixing.  


\renewcommand{\refname}{Supplementary References}
\putbib[DiamondOMCs]

\end{bibunit}


\end{document}